\documentclass[12pt]{article}
\usepackage{amssymb}
\usepackage{amsmath}
\usepackage{graphicx}

\catcode`@=11
\renewcommand{\section}{\@startsection{section}{1}{0pt}{\medskipamount}
{\medskipamount}{\large\bf}}
\numberwithin{equation}{section}
\catcode`@=12
\def\beq{\begin{eqnarray}}    %%%  begequation/eqnarray
\def\eeq{\end{eqnarray}}      %%%  endequation/eqnarray

\def\ln{\,\mbox{ln}\,}                  %%% log
\def\={\ =\ }

\begin{document}

\begin{titlepage}
\setcounter{page}{0}

\vskip 2.0cm

\begin{center}

{\large\sc Composite operators in functional renormalization group
approach}
\vspace{10mm}

{\large Olga V. Zyryanova $^{a}$}

\vspace{6mm}

\noindent ${}^{(a)}${\em
Tomsk State Pedagogical University,\\
Kievskaya St.\ 60, 634061 Tomsk, Russia}
%\vspace{4mm}

\vskip 12mm

\begin{abstract}
\noindent The gauge dependence problem of the effective  action
 for general gauge theories in the framework of a modified
functional renormalization group approach proposed recently
 is studied. It is shown that  the effective action remains gauge-dependent on-shell.

\vskip 50mm

\noindent {\sl Keywords:} \  \ Effective
action, general gauge theories, \, gauge dependence, composite
operators. \vskip 2mm

\noindent {\sl PACS:} \
%11.10.Gh, \ %% Renormalization
%11.10.Hi, \ %% Renormalization group evolution of parameters
11.15.-q, \ %% Gauge field theories
11.15.Bt \ %% General properties of perturbation theory
%11.15.Tk  \ %% Other nonperturbative techniques
\vskip 2mm

\end{abstract}
\end{center}
\hfill

\noindent{\sl E-mail:}    zyryanova@tspu.edu.ru

\end{titlepage}

%%%%%%%%%%%%%%%%%%%%%%%%%%%%%%%%%%%%%%%%%%%%%%%%%%%%%%%%%%%%%%
\section{Introduction}

\noindent

The functional renormalization group approach
\cite{Wett-1,Wett-2} is an attempt to give a method to
consider some non-perturbative aspects of quantum theories. There is
a rather high activity in applications of this method as to various
problems in the quantum field theory (see, for example,
\cite{Gies,LSh} and references therein). Special attention is given
to the gauge theories as the ones  in terms of which   all
fundamental interactions might  be described. Recently, it has been
shown  that the  application of the functional renormalization group
approach to the  Yang-Mills theories faces  the gauge dependence
problem as the  effective action being the main object of the method
and satisfying the flow equation remains gauge dependent even
on-shell \cite{LSh}. Moreover, the gauge dependence of the effective
action has been explicitly demonstrated by explicit calculations for
a simple Abelian model in linear gauges \cite{LM}. These facts tell
us that there is no possibility for a reasonable physical
interpretation of any results obtained within  the framework of
standard FRG \cite{Wett-1,Wett-2} being applied as to gauge systems.

Just recently,  a modification of the standard FRG approach has been
proposed \cite{modFRG}. The modification consists of a single
insertion of  composite operators into   the generating functional
of Green functions. Application of this proposal to the gauge
theories requires the general study of the gauge dependence problem.
In the present paper, we analyze this problem as for the general
gauge theories formulated in the field-antifield quantization
formalism, in arbitrary admissible gauges \cite{BV,BV1}. We find
that the modified FRG approach \cite{modFRG} suffers with the same
problem as the standard one \cite{Wett-1,Wett-2}.

We use the condensed DeWitt's notations \cite{DeWitt}.
The Grassmann parity of a quantity $F$ is denoted as $\varepsilon (F)$.
\\

%%%%%%%%%%%%%%%%%%%%%%%%%%%%%%%%%%%%%%%%%%%%%%%%%%%%%%%%%%%%%%%%%%%%
\section{Gauge dependence problem in the FRG approach}

Our starting point is a gauge theory described within the framework of the
field-antifield formalism \cite{BV,BV1} by the effective action
functional $S_{eff}(\varphi)$ as defined by
\beq
\label{CFa1}
S_{eff}(\varphi)=S_{ext}(\varphi,\varphi^*)\big|_{\varphi^*=0}=
S\Big(\varphi,\varphi^*+\frac{\delta\Psi}{\delta\varphi}\Big)\Big|_{\varphi^*=0},
\eeq where $\varphi=\{\varphi^A\}$
($\varepsilon(\varphi^A)=\varepsilon_A$) means the fields of the
total configuration space, including the initial fields of the
classical gauge theory, the ghost and the antighost fields, the
Nakanishi - Lautrup  fields, and so on.
%In (\ref{CFa1})
Here $\varphi^*=\{\varphi^*_A\}$
($\varepsilon(\varphi^*_A)=\varepsilon_A+1$) denotes the set of the respective
 antifields.
%to fields $\varphi$ and
Both the actions $S_{ext}=S_{ext}(\varphi,\varphi^*)$ and
$S=S(\varphi,\varphi^*)$ satisfy the quantum master equation
\beq
\label{CFa2}
\Delta\exp\Big\{\frac{i}{\hbar}S_{ext}\Big\}=0,\quad
\Delta\exp\Big\{\frac{i}{\hbar}S\Big\}=0,\quad \Delta =
(-1)^{\varepsilon_A}\frac{\overrightarrow{\delta}}{\delta\varphi^A}
\frac{\overrightarrow{\delta}}{\delta\varphi^*_A},\quad \Delta^2=0 .
\eeq
Fermion functional $\Psi=\Psi(\varphi)$ describes the gauge
fixing  characteristic to  the field-antifield formalism. In turn,
the natural arbitraries in $S_{ext}$ is described by the formula
(see, for instance, \cite{BB,BLT-15})
\beq
\label{CFa3}
\exp\Big\{\frac{i}{\hbar}S_{ext}\Big\}=
\exp\{[\Delta,\Psi]\}\exp\Big\{\frac{i}{\hbar}S\Big\},
\eeq
where
$[\;,\;]$ stands for the supercommutator,
$[F,G]=FG-GF(-1)^{\varepsilon(F)\varepsilon(G)}$.

The generating functional of the
Green functions
for the general gauge theories within the FRG approach can be written down
in the form of a functional integral over the fields $\varphi$,
\beq
\label{CFa4}
Z_k(J,K)&=&\int
D\varphi\exp\Big\{\frac{i}{\hbar}\left[S_{eff}(\varphi)+
J_A\varphi^A+S_k(\varphi)+K{\cal O}(\varphi)\right]\Big\},
\eeq
${\cal O}(\varphi)$ is an arbitrary composite operator,
$J=\{J_A\},$ $ (\varepsilon(J_A)=\varepsilon_A)$
are the respective sources to the fields $\varphi=\{\varphi^A\}$,  $K$ is the source
to the composite operator
${\cal O}(\varphi)$ $(\varepsilon(K)=\varepsilon({\cal O}(\varphi)))$.
The action $S_k(\varphi)$ named  the regulator action within
the FRG approach \cite{Wett-1,Wett-2, Gies}
 has  the quadratic form,
\beq
\label{CFa5}
S_k(\varphi)=\frac{1}{2}R_{kAB}\varphi^B\varphi^A
\equiv \int dxdy \frac{1}{2}R_{kAB}(x,y)\varphi^B(x)\varphi^A(y),
\eeq
where $R_{kAB}(x,y)$, $R_{kAB}(x,y)=R_{kBA}(y,x)(-1)^{\varepsilon_A\varepsilon_B}$ are
the so-called regulator functions. Index $k$ denotes a momentum rescaling   parameter
%in the standard FRG,
such that
\beq
\label{CFa5a}
\lim_{k\rightarrow 0}S_k(\varphi)=0.
\eeq
The standard FRG approach \cite{Wett-1,Wett-2} corresponds to the case when $K=0$
while the modified one \cite{modFRG} works with $K\neq 0$.  For both these approaches
the gauge dependence problem  is studied in a similar way.

To use the most efficient aspects  of the field-antifield formalism it is convenient
to introduce the
extended generating functional of the Green functions, $Z_k(J,\varphi^*,K)$,
\beq
\label{CFb1}
%\nonumber
Z_k(J,\varphi^*,K)=\int D\varphi
\exp\Big\{\frac{i}{\hbar}\left[S_{ext}(\varphi,\varphi^*)+
S_k(\varphi)+J_A\varphi^A+K{\cal O}(\varphi)\right]\Big\}. \eeq
Obviously, one has,
\beq
\label{CFb2}
Z_k(J,\varphi^*,K)\big|_{\varphi^*=0}=Z_k(J,K), \eeq
where
$Z_k(J,K)$  is defined in (\ref{CFa4}).

The modified Ward identity for the generating functional $Z_k=Z_k(J,\varphi^*,K)$
has the form
\beq
\label{CFb3}
J_A\frac{\overrightarrow{\delta}Z_k}{\delta\varphi^*_A}+
\frac{\hbar}{i}R_{kAB}
\frac{\overrightarrow{\delta}^2Z_k}{\delta J_B\delta \varphi^*_A}+
K\widehat{{\cal O}}_A
\frac{\overrightarrow{\delta}Z_k}{\delta\varphi^*_A}=0,
\eeq
which  follows from the equality
\beq
\label{CFb4}
\int D\varphi\exp\Big\{\frac{i}{\hbar}\left[S_k(\varphi)+
J_A\varphi^A+K{\cal O}(\varphi)\right]\Big\}\Delta
\exp\Big\{\frac{i}{\hbar}S_{ext}(\varphi,\varphi^*)\Big\}=0 .
\eeq
In (\ref{CFb3}), the notations
\beq
\label{CFb5}
\widehat{{\cal O}}_A=
{\cal O}_A\Big(\frac{\hbar}{i}\frac{\overrightarrow{\delta}}{\delta J}\Big),
\quad
{\cal O}_A(\varphi)={\cal O}(\varphi)\frac{\overleftarrow{\delta}}{\delta\varphi^A},
\eeq
are used. When $k\rightarrow 0$ and $K=0$ the (\ref{CFb3})  reduces
to the standard Ward identity playing the crucial role as to study the
renormalization and the gauge dependence problem for general gauge theories
within  the field-antifield formalism \cite{VLT-82}.

Consider now  an infinitesimal variation of the gauge fixing functional
$\Psi$ entering  the action $S_{ext}$ (\ref{CFa3})
\beq
\label{CFc1}
\Psi(\varphi)\;\rightarrow\;\Psi(\varphi)+\delta\Psi(\varphi) .
\eeq
The latter  generates  the following variation in $S_{ext}$
\beq
\label{CFc2}
\delta\exp\Big\{\frac{i}{\hbar}S_{ext}\Big\}=
\Delta \delta\Psi\exp\Big\{\frac{i}{\hbar} S_{ext}\Big\},
\eeq
and,  thereby,  in
%the infinitesimal variation of
the generating functional of the Green
functions (\ref{CFb1}), $Z_k=Z_k(J,\varphi^*,K)$,
\beq
\label{CFc3}
\delta Z_k=\int D\varphi\exp\Big\{\frac{i}{\hbar}\left[S_k(\varphi)+
J_A\varphi^A+K{\cal O}(\varphi)\right]\Big\}
\Delta\delta\Psi(\varphi)
\exp\Big\{\frac{i}{\hbar}S_{ext}(\varphi,\varphi^*)\Big\}.
\eeq
By usual manipulations, the equation (\ref{CFc3}) rewrites
in the form
\beq
\label{CFc4}
\delta Z_k=-\frac{i}{\hbar}\Big(J_A+\frac{\hbar}{i}R_{kAB}
\frac{\overrightarrow{\delta}}{\delta J_B}+K\widehat{{\cal O}}_A\Big)
\delta\widehat{\Psi}(-1)^{\varepsilon_A}
\frac{\overrightarrow{\delta}Z_k}{\delta\varphi^*_A},
\eeq
where $\widehat{{\cal O}}_A$ is defined in (\ref{CFb5}) and
\beq
\label{CFc5}
\delta\widehat{\Psi}=\delta \Psi\Big(\frac{\hbar}{i}
\frac{\overrightarrow{\delta}}{\delta J}\Big).
\eeq
By using the Ward identity (\ref{CFb3}),  the equation defining
the gauge dependence of the effective action
(\ref{CFc4}) rewrites in a very nice form
\beq
\label{CFc6}
\delta Z_k=\frac{i}{\hbar}[\delta\widehat{\Psi},J_A]
\frac{\overrightarrow{\delta}Z_k}{\delta\varphi^*_A}=
\delta\widehat{\Psi}_A\frac{\overrightarrow{\delta}Z_k}{\delta\varphi^*_A},
\eeq
where
\beq
\label{CFc7}
\delta\widehat{\Psi}_A=\delta\Psi_A
\Big(\frac{\hbar}{i}\frac{\overrightarrow{\delta}}{\delta J}\Big),\quad
\delta\Psi_A(\varphi)=\delta\Psi(\varphi)
\frac{\overleftarrow{\delta}}{\delta\varphi^A}.
\eeq
It follows from (\ref{CFc4}) that the relation holds
\beq
\label{CFc8}
\delta\widehat{\Psi}_A
\frac{\overrightarrow{\delta}Z_k}{\delta\varphi^*_A}\Big|_{J=0,K=0}\neq 0 .
\eeq
The relation (\ref{CFc8}) allows one  to reveal
 the gauge dependence of the average effective
action in the both FRG approaches, even on-shell.
This means that there is no consistent  physical description of the results
obtained within the framework
of the standard  FRG approach \cite{Wett-1,Wett-2}, or of the modified one \cite{modFRG},
in the case of gauge theories.
\\

\section{Discussion}

Notice that there exists a way  to solve the gauge dependence  problem appearing
in the functional renormalization group approach with use the concept of  quantum
field theory with composite operators \cite{DeDM,CJT,L}.
Following the paper \cite{LSh}
one can extend the
action  in the functional integral (\ref{CFa4}) with the new term,
$K_{1} L_{k}( \phi )$ instead of $S_k(\varphi)$, introducing the
new external source $K_{1}$ as for the respective quantity
$L_{k}( \phi )$. Then one can state that the effective action
with these composite operators does not depend on gauges on-shell which
is defined with respect to the equations of motion as for the effective action.

Quite recently  the problem of modification of the standard quantization rules
not destroying the gauge status of the scheme was discussed in the paper \cite{BL}.
It was proven that defining the
physical observable quantities, $\mathcal{ O }_{ phys }$,  in the
usual way, one extends the action in the functional integral
with the new term $\frac{ \hbar }{ i } \ln \mathcal{ O }_{ phys }
( \phi )$. There is no new external sources  introduced in the
latter case.  It follows that the standard quantum master equation
absorbs consistently the new term, provided that the $\mathcal{ O }_
{ phys }$ is annihilated by the BRST operator $\sigma  =:  \frac{ \hbar }
{ i } \Delta  +  {\rm ad}( S_{eff} )$, where ${\rm ad}( S_{eff} )$ means the adjoint
antibracket operator.  Then  it follows that
the total effective action is gauge independent on-shell. From this point of view
the gauge dependence problem in the standard FRG \cite{Wett-1,Wett-2}
is related with the fact that $\sigma\exp\{\frac{i}{\hbar}S_k\}\neq 0$.
\\

\section*{Acknowledgments}
\noindent
The author thanks P.M. Lavrov for useful discussions.
The work  is supported by the Ministry of Education and Science of
Russian Federation, grant  3.1386.2017 and by the RFBR grant
15-02-03594.


\begin {thebibliography}{99}
\addtolength{\itemsep}{-8pt}


\bibitem{Wett-1} C. Wetterich,
{\it Average Action And The Renormalization Group Equations.}
Nucl. Phys. B352 (1991) 529.

\bibitem{Wett-2} C. Wetterich,
{\it Exact evolution equation for the effective potential,}
Phys. Lett. B301 (1993) 90. %% –94

\bibitem{Gies} H. Gies,
{\it Introduction to the functional RG and applications to
gauge theories},
Lect. Notes Phys. 852 (2012) 287.%-348; hep-th/0611146.

\bibitem{LSh}
P. M. Lavrov, I. L. Shapiro, {\it On the functional
renormalization group approach for Yang-Mills fields},
JHEP {\bf 06} (2013) 086.

\bibitem{LM}
P. M. Lavrov, B. S. Merzlikin, {\it Loop expansion of average
effective action in functional renormalization group approach},
Phys. Rev. D {\bf 92} (2015) 085038.

\bibitem{modFRG}
C. Pagani, M. Reuter, {\it Composite Operators in Asymptotic Safety},
arXiv:1611.06522 [gr-qc].

\bibitem{BV}
I. A. Batalin, G. A. Vilkovisky, {\it Gauge algebra and quantization},
Phys. Lett. B {\bf 102} (1981) 27.% - 31.

\bibitem{BV1}
I. A. Batalin, G. A. Vilkovisky, {\it Quantization of gauge theories with linearly
dependent generators}, Phys. Rev. D {\bf 28} (1983) 2567.% - 2582.

\bibitem{DeWitt}
B. S. DeWitt, {\it Dynamical Theory of Groups and Fields},
Gordon and Breach, New York, 1965.

\bibitem{BB}
I. A. Batalin, K. Bering,
{\it Gauge Independence in a Higher-Order Lagrangian
Formalism via Change of Variables in the Path Integral},
Phys. Lett. B {\bf 742} (2015) 23.%-28

\bibitem{BLT-15}
I. A. Batalin, P. M. Lavrov, I. V. Tyutin,
{\it Finite anticanonical transformations in field–antifield formalism},
Eur. Phys. J. C {\bf 75} (2015) 270.

\bibitem{VLT-82} B. L. Voronov, P. M. Lavrov, I. V. Tyutin,
{\it Canonical transformations and the gauge dependence
in general gauge theories,} Sov. J. Nucl. Phys. {\bf 36} (1982) 292.

\bibitem{DeDM}
C. De Dominicis, P. C. Martin, {\it Stationary entropy principle and renormalization in
normal and superfield systems. I. Algebraic formulation},
J. Math. Phys. {\bf 5} (1964) 14.%-30.

\bibitem{CJT} J. M. Cornwall, R. Jackiw, E. Tomboulis,
{\it Effective action for composite operators},
Phys. Rev. D {\bf 10} (1974) 2428.

\bibitem{L}
P. M. Lavrov,
{\it Effective action for composite fields in gauge theories},
Theor. Math. Phys. {\bf 82} (1990) 282.

\bibitem{BL}
I. A. Batalin, P. M. Lavrov, {\it Physical quantities  and arbitrariness
in resolving quantum master equation}, arXiv:1702.02663 [hep-th].

\end{thebibliography}
\end{document}